\documentclass[reprint,aps,prl,twocolumn,superscriptaddress,longbibliography]{revtex4-2}
\usepackage{amsmath,amssymb,bm,braket}\usepackage{tabularx,multirow,booktabs,dcolumn} 
\usepackage[table,xcdraw]{xcolor} \definecolor{gray}{rgb}{0.5,0.5,0.5}

\usepackage[normalem]{ulem}\usepackage{graphicx}\usepackage[colorlinks=true,linkcolor=blue,citecolor=blue,urlcolor=blue]{hyperref}

\begin{document} 

\title{Large Anomalous Hall Effect in Topologically Trivial Double-$Q$ Magnets}
\author{Satoru Ohgata}\author{Satoru Hayami} 
\affiliation{Graduate School of Science, Hokkaido University, Sapporo 060-0810, Japan.}

\begin{abstract}
    Multi-$Q$ magnets consist of superposed spin density waves with distinct magnetic modulation vectors, enabling a wide range of magnetic orders depending on their combination. 
    Among them, topologically nontrivial spin textures, such as a magnetic skyrmion, has been extensively studied owing to the emergence of topological Hall effects induced by real-space scalar spin chirality. 
    Contrary to this expectation, we theoretically investigate another route to enhancing the Hall response under a topologically \textit{trivial} double-$Q$ spin textures.
    Despite the cancellation of the scalar spin chirality, the double-$Q$ magnetism exhibits a pronounced Hall response with a nonmonotonic dependence on the uniform magnetization, which is in stark contrast to a ferromagnetic state and a single-$Q$ spiral state. 
    Analyzing the multi-orbital Kondo lattice model, we show that orbital hybridization induced by the double-$Q$ superstructure enhances the Berry curvature in $\bm{k}$-space, leading to a large anomalous Hall effect.
    This mechanism accounts for the observed giant anomalous Hall effect in GdRu$_2$Si$_2$ and GdRu$_2$Ge$_2$, thereby highlighting topologically trivial double-$Q$ spin textures as promising spintronic materials. 
\end{abstract}

\maketitle

Multi-$Q$ magnets are systems in which multiple spin density waves (SDWs) with distinct magnetic modulation vectors $\bm{Q}$ are superposed~\cite{Bak_PhysRevLett.40.800, Izyumov_1984_MultiQ, McEwen_PhysRevB.34.1781}. 
Depending on the combinatoin of SDWs involved, multi-$Q$ magnets can realize a wide range of magnetic orders~\cite{Binz_PhysRevLett.96.207202, Binz_PhysRevB.74.214408, nagaosa2013topological, GOBEL_2021_skyrmion_review, hayami2024stabilization}. 
Among these, noncoplanar spin textures are of particular interest due to their finite scalar spin chirality in real space~\cite{Holstein_1968_THE,Bruno_2004_THE}.
This scalar spin chirality acts as an emergent magnetic field, which induces a transverse current even without an external magnetic field or relativistic spin--orbit coupling (SOC), i.e., the topological Hall effect (THE)~\cite{Loss_PhysRevB.45.13544, Ye_PhysRevLett.83.3737, Ohgushi_2000, Shindou_2001,Tatara_2002,Neubauer_2009, Kanazawa_2011,Hamamoto_2015,Gobel_2017_skyrmion, Kurumaji_2019}.
In particular, topologically nontrivial spin textures—such as skyrmions—have attracted considerable attention~\cite{Muhlbauer_2009_skyrmion,yu2010real, Romming_2013_skyrmion, Fert_2013_skyrmion, Fert_2017_skyrmion, Gobel_2017_skyrmion}, since the skyrmion number 
\begin{equation}
    N_{\mathrm{sk}} = \frac{1}{4\pi} \iint \bm{n}(\bm{r}) \cdot \left( \partial_x \bm{n}(\bm{r}) \times \partial_y \bm{n}(\bm{r}) \right)\, dx\, dy
\end{equation}
is directly proportional to the magnitude of the THE~\cite{Verma_2022_THE_AHE_relation}.

In contrast, the mechanisms responsible for the enhanced anomalous Hall effect (AHE) in topologically trivial spin textures (i.e., without net scalar spin chirality) are not yet well elucidated~\cite{Nagaosa_RevModPhys.82.1539, Xiao_RevModPhys.82.1959}.
Moreover, the AHE in such topologically trivial magnets has generally been assumed to vary monotonically with magnetic field or magnetization. 
However, recent experiments have revealed that even topologically trivial double-$Q$ (2$Q$) magnetic structures with $N_{\mathrm{sk}}=0$ can exhibit a large AHE comparable to that of skyrmion textures, with a nonmonotonic magnetization dependence~\cite{Khanh_2020_GdRu2Si2, Yoshimochi_2024_GdRu2Ge2}.
These observations challenge the conventional expectation that topologically trivial spin textures cannot yield a large AHE, indicating the presence of an alternative enhancement mechanism.

In this study, we theoretically propose such an alternative route to enhancing the AHE without relying on topologically nontrivial spin textures, and show that topologically trivial 2$Q$ magnets serve as prototypical examples exhibiting a giant AHE.
We perform theoretical calculations based on the multi-orbital Kondo lattice model~\cite{Kasuya_1956_Zener_model, Kondo_1964_Resistance_Minimum, Tsunetsugu_1997_Kondo_model_review}, and find that the 2$Q$ state develops a large Berry curvature in $\bm{k}$-space, despite the absence of net scalar spin chirality in real space.
We further demonstrate that orbital hybridization caused by the 2$Q$ superstructure is the microscopic origin of 
such enhancement, leading to behavior that is qualitatively distinct from both ferromagnetic and single-$Q$ (1$Q$) states. 
Furthermore, the AHE in the 2$Q$ state exhibits a nonmonotonic dependence on the magnetic field or magnetization, in agreement with the experimental observations in GdRu$_2$Si$_2$~\cite{Khanh_2020_GdRu2Si2} and GdRu$_2$Ge$_2$~\cite{Yoshimochi_2024_GdRu2Ge2}.
Our results thus uncover a fundamental mechanism by which topologically \textit{trivial} spin textures can host a giant AHE.

\begin{figure*}[t]
    \centering
    \includegraphics[width=\textwidth]{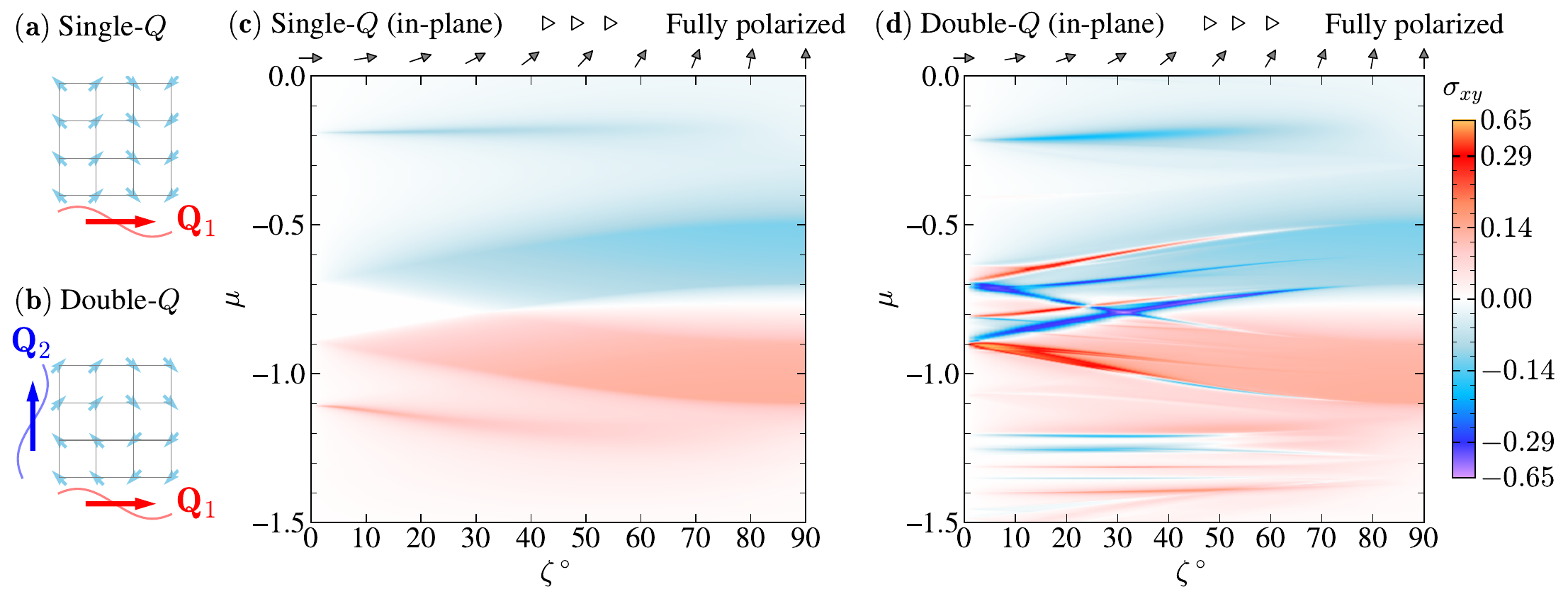}
    \caption{
    Panels (a) and (b) illustrate the 1$Q$ and 2$Q$ spin configurations, respectively. 
    Panels (c) and (d) show the anomalous Hall conductivity $\sigma_{xy}$ for the 1$Q$ and 2$Q$ states as functions of the chemical potential $\mu$ and the out-of-plane canting angle $\zeta$ of the magnetic moments. 
    $\zeta = 0^{\circ}$ corresponds to coplanar spin configurations, while $\zeta = 90^{\circ}$ represents a fully polarized configuration.
    The same color scale is used for panels (c) and (d).
    For $-1.5 < \mu < 0$, the maximum values of $|\sigma_{xy}|$ are 0.647 for the 2$Q$ state and 0.145 for the 1$Q$ state.
    The same figure, plotted over the chemical-potential range $0.5 < \mu < 2.5$, is shown in Fig.~1 of the Supplemental Materials~\cite{Suppl}.
    }\label{fig:SIGMA_heatmap}
\end{figure*}

We consider the multi-orbital Kondo lattice model~\cite{Kasuya_1956_Zener_model, Kondo_1964_Resistance_Minimum, Tsunetsugu_1997_Kondo_model_review}, whose Hamiltonian is given by
\begin{align}\label{eq:H_total}
    \mathcal{H} 
    &= \sum_{i,i'\ell,\ell^{\prime},s}        
    t_{i i^{\prime}}^{\ell \ell^{\prime}}\,  c_{i \ell s}^{\dagger} c_{i^{\prime} \ell^{\prime}s} 
   + \lambda \sum_{i} (l_z\sigma_z)_{i}
   + J \sum_{i} \bm{\sigma}_{i} \cdot \bm{S}_i,
\end{align}
where $c_{i \ell s}^{\dagger}$ ($c_{i \ell s}$) creates (annihilates) an electron at atomic site $i$, with orbital $\ell \in \{p_x, p_y\}$ and spin $s \in \{\uparrow, \downarrow\}$, on a square lattice.
The first term represents the hopping Hamiltonian, where $t_{i i^{\prime}}^{\ell \ell^{\prime}}$ denotes the matrix element from orbital $\ell^{\prime}$ at site ${i^\prime}$ to orbital $\ell$ at site $i$.
We adopt the Slater-Koster parameters $t_{pp\sigma}$ and $t_{pp\pi}$ for the nearest-neighbor hopping, and $t_{pp\sigma}^{\prime}$ and $t_{pp\pi}^{\prime}$ for the next-nearest-neighbor hopping.
The second term describes the atomic SOC in the $\{p_x, p_y\}$, representing the coupling between orbital angular momentum $l_z$ and spin $\sigma_z$: $(l_z \sigma_z)_{i} = \sum_{\ell,\ell^{\prime},s,s^{\prime}} \langle \ell | l_z | \ell^{\prime} \rangle \langle s | \sigma_z | s^{\prime} \rangle c_{i \ell s}^{\dagger} c_{i \ell^{\prime} s^{\prime}}$; $\lambda$ is the coupling constant.
The third term represents the Kondo coupling with the coupling constant $J$; $\bm{S}_i$ denotes the classical localized spin at site $i$ with $|\bm{S}_i| = 1$.
$\bm{\sigma}_{i} = \sum_{\ell,s,s^{\prime}} \langle s \vert \bm{\sigma} \vert s^{\prime} \rangle\, c_{i \ell s}^{\dagger} c_{i \ell s^{\prime}}$ with the vector of Pauli matrices $\bm{\sigma} = (\sigma^x, \sigma^y, \sigma^z)$ denotes the itinerant electron spin.

For the localized spin configurations $\{\bm{S}_i\}$, we consider two types of magnetic states. 
The first is the 1$Q$ state with the modulation vector $\bm{Q}_1=\pi/2\;\bm{e}_x$ given by
\begin{align}\label{eq:spin_config_1Q}
    \bm{S}^{\rm 1Q}_i  = (\cos \zeta \sin \mathcal{Q}_{1i},\;\; \cos \zeta \cos \mathcal{Q}_{1i},\;\; \sin \zeta),
\end{align} 
and the second is the 2$Q$ state composed of a sinusoidal superposition of $\bm{Q}_1$ and $\bm{Q}_2=\pi/2\;\bm{e}_y$, expressed as 
\begin{align}\label{eq:spin_config_2Q}
    \bm{S}^{\rm 2Q}_i  = (-\cos \zeta \cos \mathcal{Q}_{2i},\;\; \cos \zeta \cos \mathcal{Q}_{1i},\;\; \sin \zeta),
\end{align}
where $\mathcal{Q}_{\nu i}=\bm{Q}_\nu \cdot (\bm{r}_{i}-\bm{\delta})$ for $\nu=1,2$, $\bm{r}_i$ denotes the position vector, $\bm{\delta} = (\bm{e}_x+\bm{e}_y)/2$ is the phase factor, and $0^{\circ}\leq \zeta \leq 90^{\circ}$ represents the canting angle. 
At $\zeta=0^{\circ}$, $\bm{S}^{\rm 1Q}_i$ and $\bm{S}^{\rm 2Q}_i$ correspond to the 1$Q$ and 2$Q$ spin textures in the $xy$-plane, as illustrated in Figs.~\ref{fig:SIGMA_heatmap}(a) and \ref{fig:SIGMA_heatmap}(b), respectively. 
Increasing $\zeta$ cants the spins toward the $z$-direction, leading to noncoplanar spin configurations. 
Although the scalar spin chirality, $\bm{S}_i \cdot (\bm{S}_j\times \bm{S}_k)$, is locally induced in both states, it vanishes upon spatial averaging, i.e., $N_{\rm sk}=0$, indicating that the 1$Q$ and 2$Q$ states are topologically trivial. 
At $\zeta = 90^{\circ}$, both configurations continuously evolve into the fully polarized (FP) ferromagnetic state with $\bm{S}_i =\bm{e}_z$.  
The 1$Q$ state is stabilized by the Ruderman-Kittel-Kasuya-Yosida interaction~\cite{Ruderman, Kasuya, Yosida1957, yoshimori1959new, Kaplan_PhysRev.124.329, Elliott_PhysRev.124.346}, while the 2$Q$ state is stabilized by additionally taking into account the higher-order multi-spin interactions~\cite{Hayami_PhysRevB.95.224424} and magnetic anisotropy~\cite{Hayami_doi:10.7566/JPSJ.89.103702, Utesov_PhysRevB.103.064414, Wang_PhysRevB.103.104408}. 
Especially, a noncoplanar 2$Q$ state with a nonzero $\zeta$ has been identified in the skyrmion-hosting materials, such as GdRu$_2$Si$_2$~\cite{Khanh_2020_GdRu2Si2} and GdRu$_2$Ge$_2$~\cite{Yoshimochi_2024_GdRu2Ge2}.

We calculate the anomalous Hall conductivity based on the linear response theory, which is given by
\begin{equation}\label{eq:sigma_xy} 
    \sigma_{xy} = - \frac{e^2}{\hbar} \frac{1}{N} \sum_{n\bm{k}} f(\mu, E_{n\bm{k}}) \Omega_{xy}^{n\bm{k}},
\end{equation} 
where $e$ is the elementary charge, $\hbar$ is the reduced Planck constant, and $N$ denotes the total number of sites. 
$f(\mu, E_{n\bm{k}}) = [1 + \exp((E_{n\bm{k}} - \mu)/(k_{\mathrm{B}}T))]^{-1}$ denotes the Fermi--Dirac distribution function, where $k_{\mathrm{B}}$ is the Boltzmann constant, $T$ is the temperature, and $\mu$ is the chemical potential.
$\Omega_{xy}^{n\bm{k}}$ denotes the Berry curvature of the $n$-th band, calculated as
\begin{align} \label{eq:Omega_nk} 
    \Omega_{xy}^{n\bm{k}} &= -2 \hbar^2 \sum_{m \neq n} 
    \frac{ \Im \!\left[
        \langle u_{m\bm{k}} | \hat{v}_{x}^{\bm{k}} | u_{n\bm{k}} \rangle 
        \langle u_{n\bm{k}} | \hat{v}_{y}^{\bm{k}} | u_{m\bm{k}} \rangle 
    \right] }{ (E_{n\bm{k}} - E_{m\bm{k}})^2 } .
\end{align}
Here, $\hat{\bm{v}} \equiv \partial \hat{\mathcal{H}} / \partial (\hbar \bm{k})$ is the velocity operator, and $u_{n\bm{k}}$ is the periodic part of the Bloch wave function.
In the following, the $\bm{k}$ sum is evaluated on a grid of $\Delta k_x = \Delta k_y = \pi/1024$, the temperature is set to zero, and $\hbar=e=k_{\rm B}=1$.
We evaluate $\sigma_{xy}$ using the fixed spin configurations in Eqs.~(\ref{eq:spin_config_1Q}) and (\ref{eq:spin_config_2Q}), a standard approach to examine its behavior~\cite{Hamamoto_2015, Wang_PhysRevB.103.104408}.
We set the model parameters as $t_{pp\sigma}=1$, $t_{pp\pi}=-0.4$, $t_{pp\sigma}^{\prime}=0.3$, $t_{pp\pi}^{\prime}=-0.1$, and $\lambda=0.05$, and consider the weak-coupling regime by setting $J = 0.2$ unless otherwise noted; the results for other hopping parameters are shown in Supplemental Materials~\cite{Suppl}.

Figures~\ref{fig:SIGMA_heatmap}(c) and \ref{fig:SIGMA_heatmap}(d) show the calculated results of AHE for the 1$Q$ and 2$Q$ systems, respectively, where the chemical potential is sampled with a uniform spacing of $\Delta\mu = 2\times10^{-4}$ and the canting angle is sampled with a spacing of $\Delta\zeta = 1^\circ$.
For comparison, we use the same color scale in both plots.
The maximum values of $|\sigma_{xy}|$ are given by 0.647 for the 2$Q$ state and 0.145 for the 1$Q$ state. 
This demonstrates a substantial enhancement of the Hall conductivity in the 2$Q$ state compared with the 1$Q$ state.

Before discussing the AHE in the 1$Q$ and 2$Q$ states in detail, we first examine its behavior in the FP state realized at $\zeta=90^{\circ}$ in both panels [Fig.~\ref{fig:SIGMA_heatmap}(c) and (d)]. 
In this case, $\sigma_{xy}$ exhibits two broad peaks at $\mu=-1, -0.6$, which correspond to the exchange energy splitting $2J=0.4$, while their widths are primarily determined by the SOC $2\lambda=0.1$. 
This behavior is fully consistent with the conventional mechanism~\cite{Nagaosa_RevModPhys.82.1539, Xiao_RevModPhys.82.1959}, in which the interplay between SOC and ferromagnetic moments generates Berry curvatures of opposite sign for a split Kramers pair~\cite{Ohgata_2025_AHE_by_AMD}.

Next, we examine the case of $\zeta < 90^{\circ}$ in the 1$Q$ state. 
As shown in Fig.~\ref{fig:SIGMA_heatmap}(c), the behavior of $\sigma_{xy}$ is not significantly different from that in the ferromagnetic state at $\zeta=90^{\circ}$: the two main peaks at $\mu=-1$ and $\mu=-0.6$ move closer together with decreasing $\zeta$, reflecting the reduction of the $z$-spin component (magnetization). 
Meanwhile, additional structures appear around $\mu=-1.1$ and $\mu=-0.2$ at small $\zeta$, which originate from the 1$Q$ modulations. 
Their magnitudes are comparable to the ferromagnetic contributions, but they gradually diminish as $\zeta$ increases. 
Thus, the overall behavior remains qualitatively similar to that of the FP state, indicating that it is primarily governed by the interplay between SOC and the net magnetization, although the 1$Q$ spiral modulation introduces additional features in $\sigma_{xy}$ for small $\zeta$.

Compared with the 1$Q$ state in Fig.~\ref{fig:SIGMA_heatmap}(c), the behavior of $\sigma_{xy}$ changes drastically in the 2$Q$ state, as shown in Fig.~\ref{fig:SIGMA_heatmap}(d).
In addition to the features inherited from the 1$Q$ state, the 2$Q$ state exhibits additional peak structures with pronounced enhancements (up to roughly five times those in the 1$Q$ state) and even sign reversals in several regions. 
In particular, $\sigma_{xy}$ is strongly enhanced near $\mu=-0.9$ and $-0.7$, corresponding to the band energies split by the exchange coupling, $E=-0.8 \pm J/2$. 
As detailed below, this large enhancement originates from gap opening and orbital hybridization induced through electric quadrupolar scattering mechanism under the 2$Q$ spin configuration.

\begin{figure}[t]
    \centering
    \includegraphics[width=\linewidth]{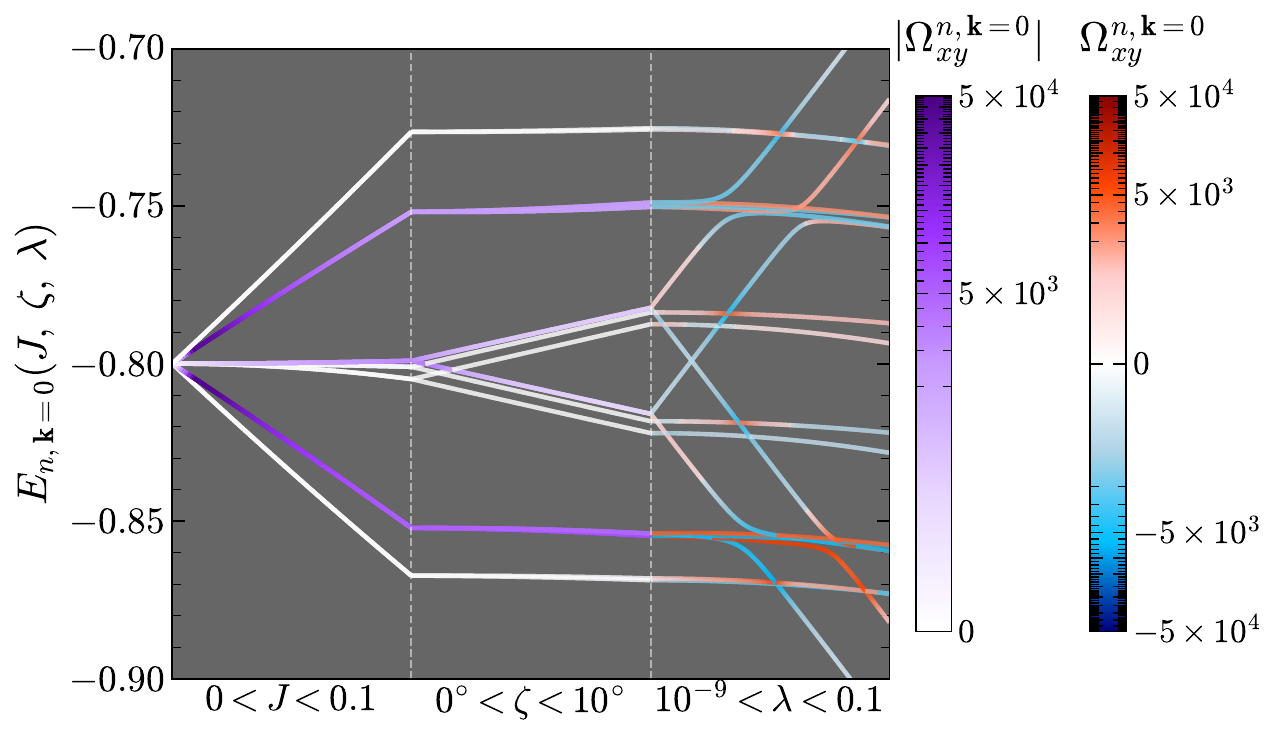}
    \caption{
    Berry curvature of the energy eigenstates at the $\Gamma$ point for the 2$Q$ magnetic configuration.
    The vertical axis denotes the eigenenergy, and the horizontal axis represents the trajectory in the model-parameter space $(J,\;\zeta,\;\lambda)$.
    Along the parameter path, only one parameter is varied at a time while the others are held fixed:
    $(J,\;\zeta,\;\lambda): (0,\;0^{\circ},\;10^{-9}) \rightarrow (0.1,\;0^{\circ},\;10^{-9}) \rightarrow (0.1,\;10^{\circ},\;10^{-9}) \rightarrow (0.1,\;10^{\circ},\;0.1)$.
    Violet shading in the region for $0<J<0.1$ and $0^{\circ}<\zeta< 10^{\circ}$ indicates the absolute value in degenerate regions where contributions cancel, 
    while the blue-to-red colormap is used in the region for $10^{-9} < \lambda < 0.1$ where degeneracy is lifted.
    }
    \label{fig:Omega_2Q}
\end{figure}

To analyze the microscopic origin of the enhanced AHE peak in the 2$Q$ state [Fig.~\ref{fig:SIGMA_heatmap}(d)], we examine the parameter dependence of the $\bm{k}$-space Berry curvature.
In particular, we focus on the Berry curvature at the $\Gamma$ point near $E=-0.8$, since the $2Q$ contribution in Fig.~\ref{fig:SIGMA_heatmap}(d) originates from the gaps opened at or near the $\Gamma$ point (see Fig.~2 of the Supplemental Material~\cite{Suppl}).
The results are shown in Fig.~\ref{fig:Omega_2Q}, where we vary three parameters $(J,\;\zeta,\;\lambda)$ in order to discuss the origin of large Berry curvature under the 2$Q$ ordering.
In the absence of the SOC, the Hall conductivity $\sigma_{xy}$ vanishes because 2$Q$ spin configuration carries no net scalar spin chirality, and each band remains at least doubly degenerate, ensuring that the Berry curvature cancels within each degenerate pair. 
Remarkably, however, a large Berry curvature of order $10^{4}$ appears as a hidden form in the doubly degenerate bands even for an almost negligible SOC $\lambda = 10^{-9}$, as indicated by the violet region in the range $0 < J < 0.1$, where the 2$Q$ spin configuration is characterized by a coplanar arrangement. 
A similar trend persists for $0^{\circ} < \zeta < 10^{\circ}$, where the system hosts a noncoplanar yet topologically trivial spin configuration.
This behavior is unique to the 2$Q$ state and does not appear in the 1$Q$ state, as shown in Supplemental Material~\cite{Suppl}.
Reflecting this distinction, the application of SOC leads to quantitatively different responses in $\sigma_{xy}$ for the two states.

The introduction of the SOC in the range for $10^{-9} < \lambda< 0.1$ lifts the band degeneracy and gives rise to pronounced values of the net Berry curvature.
The magnitude of the induced Berry curvature is comparable to that found in the degenerate bands, indicating that the giant AHE originates from the large Berry curvature already present within the degenerate manifold. 
In this sense, a large SOC is not necessary, since its primary role here is merely to lift the band degeneracy.
Moreover, the giant Berry curvature embedded within the degenerate bands is a distinctive feature of the 2$Q$ state and is absent in the 1$Q$ phase. 
This observation underscores the essential role of multi-$Q$ spin textures in generating such an enhanced Berry-curvature response.

A key distinction between the 1$Q$ and 2$Q$ states lies in the presence or absence of interference between the constituent spin density waves. 
In the 2$Q$ state, the spin texture is formed by two modulations with wave vectors $\bm{Q}_1$ and $\bm{Q}_2$, whose superposition gives rise to additional spin-scattering channels that strongly affect the conduction electrons. 
For example, multi-spin scattering processes involving both $\bm{Q}_1$ and $\bm{Q}_2$ generate quantities at $\bm{Q}_1+\bm{Q}_2$, schematically denoted as $(S^x_{\bm{Q}_1} S^y_{\bm{Q}_2})$ ($S^\mu_{\bm{q}}$ is the Fourier component of the real spin at $\mu$ component and wave vector $\bm{q}$), which correspond---via the exchange interaction $J$ between itinerant electrons and localized spins---to an electric quadrupolar scattering mechanism that hybridizes the $p_x$ and $p_y$ orbitals~\cite{Hayami_PhysRevB.104.144404}. 
Such quadrupolar scattering processes modify the orbital magnetic dipole at $\bm{Q}_1+\bm{Q}_2$, whose matrix elements are the same as the electric quadrupole, thereby inducing substantial modulations of the Berry curvature.
Indeed, as shown in the Supplemental Material~\cite{Suppl}, the behavior of the orbital magnetic dipole at $\bm{Q}_1+\bm{Q}_2$ closely tracks the field and parameter dependence of $\sigma_{xy}$ in the 2$Q$ phase. 
This correspondence indicates that higher-harmonic orbital hybridization, arising from the interference inherent to the 2$Q$ spin texture, is essential for enhancement of the AHE in this phase.

\begin{figure}[t]
    \centering
    \includegraphics[width=\linewidth]{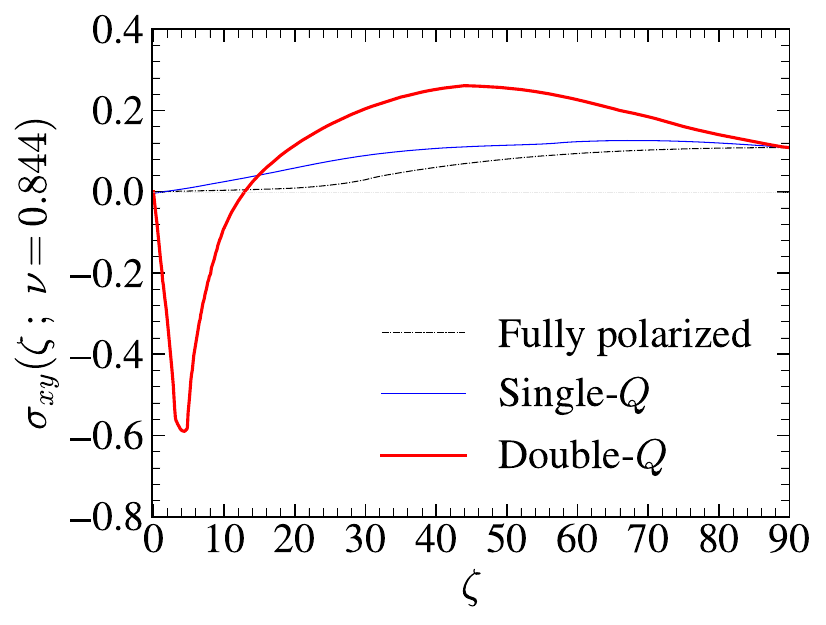}
    \caption{
    Anomalous Hall conductivity~$\sigma_{xy}$ as a function of the parameter~$\zeta$.
    The three line plots correspond to the fully polarized (black dash-dotted), 1$Q$ (blue solid), and 2$Q$ (red bold) magnetic states, respectively.
    In all cases, the electron filling ratio $\nu$ is fixed at~0.84, correspnding to $\mu \simeq 1.77$ for $\zeta=0^{\circ}$ (See Fig.~1 of the Supplemental Material~\cite{Suppl}).
    The sampling interval for the canting angle $\zeta$ is $\Delta\zeta = 1^\circ$ in all cases, except for the 2$Q$ configuration in the range $0^\circ < \zeta < 10^\circ$, for which a finer resolution of $\Delta\zeta = 0.1^\circ$ is employed to capture the rapid variation in $\sigma_{xy}$.
    }\label{fig:SIGMA_filling_const}
\end{figure}

To further elucidate the unconventional AHE arising from the 2$Q$ spin configuration, we examine the dependence of $\sigma_{xy}$ on the out-of-plane angle $\zeta$ (mimicking out-of-plane magnetic field or magnetization) at a fixed electron filling, as shown in Fig.~\ref{fig:SIGMA_filling_const}. 
For comparison, we also include the FP configuration $\bm{S}_{\mathrm{FP}}(x,y)=\sin\zeta\,\boldsymbol{e}_z$.
In both the FP state and the 1$Q$ state, $\sigma_{xy}$ increases monotonically with $\zeta$, reflecting the conventional behavior in which the AHE is governed primarily by the net magnetization and its interplay with SOC.  
The effect in the 1$Q$ state is modest, arising only from the gap openings generated by the 1$Q$ modulation.

The situation changes qualitatively in the 2$Q$ state.  
Here, $\sigma_{xy}$ shows a striking nonmonotonic evolution as a function of $\zeta$: it grows rapidly and reaches a maximum near $\zeta \approx 4^\circ$, then oscillates (changing its sign) before converging to the value of the FP state at large $\zeta$.  
This behavior stands in clear contrast to the trends in the 1$Q$ and FP states, reflecting the fact that the 2$Q$ spin texture induces a nontrivial reconstruction of the Berry curvature in $\bm{k}$-space. 
The origin of the non-monotonic response can be attributed to the Zeeman-driven shift of Berry-curvature hot spots in the electronic structure.  
As $\zeta$ increases, these hot spots move in energy, and the AHE is strongly enhanced when two such hot spots coincide at the Fermi level.

Such a non-monotonic dependence of the AHE is consistent with recent experimental observations in skyrmion-hosting materials GdRu$_2$Si$_2$ and GdRu$_2$Ge$_2$~\cite{Khanh_2020_GdRu2Si2,Khanh_2022_MultiQ_GdRu2Si2,Yasui_2020_GdRu2Si2, Matsuyama_PhysRevB.107.104421, Wood_PhysRevB.107.L180402, eremeev2023insight, Yoshimochi_2024_GdRu2Ge2, dong2025pseudogap, Huddart_PhysRevB.111.054440}. 
In both materials, enhanced AHE has been observed in the high-field region where a 2$Q$ spin configuration has been proposed to emerge~\cite{Khanh_2022_MultiQ_GdRu2Si2, Yoshimochi_2024_GdRu2Ge2}; 
$\sigma_{xy}$ exhibits a pronounced peak followed by a sign reversal or suppression as the magnetic field increases, which cannot be explained within a conventional AHE mechanism proportional to magnetization. 
The resemblance between these experimental trends and our calculated $\zeta$-dependent response strongly suggests that the non-topological 2$Q$ spin configuration itself plays an essential role in shaping the high-field AHE in these materials.

To summarize, we show that a large intrinsic AHE can arise even in a topologically trivial spin texture with 2$Q$ spin modulations. 
Employing the multi-orbital Kondo lattice model, we demonstrate that the enhanced AHE appears only in the 2$Q$ state, in clear contrast to the ferromagnetic and 1$Q$ spiral states. 
We further show that higher-order spin-scattering processes inherent to the 
2$Q$ spin texture induce effective orbital hybridization---as evidenced by the orbital angular momentum at $\bm{Q}_1+\bm{Q}_2$---which in turn produces the Berry-curvature hot spots responsible for the enhanced AHE.

We also find that the AHE induced by the 2$Q$ spin texture exhibits a nonmonotonic dependence on the magnetization, including sign reversals and sharp enhancements.
This behavior distinguishes it from the behavior of the 1$Q$ spin texture and fully polarized state, and consistent with experimental observations in GdRu$_2$Si$_2$ and GdRu$_2$Ge$_2$. 
These results highlight multi-$Q$ interference as an effective mechanism for realizing large AHEs beyond conventional topological or ferromagnetic origins.

This research was supported by JSPS KAKENHI Grants Numbers JP22H00101, JP22H01183, JP23H04869, JP23K03288, JP23K20827, and by JST CREST (JPMJCR23O4) and JST FOREST (JPMJFR2366).

\bibliographystyle{apsrev4-2}
\bibliography{ahe}

\end{document}

\title{Supplementary Material: Large Anomalous Hall Effect in Topologically Trivial Double-$Q$ Magnets}
\author{Satoru Ohgata}\author{Satoru Hayami} \affiliation{Graduate School of Science, Hokkaido University, Sapporo 060-0810, Japan.}

\maketitle
\subsection{Anomalous Hall conductivity as a function of canting angle and chemical potential}
\begin{figure}[t]
    \centering
    \includegraphics[width=\textwidth]{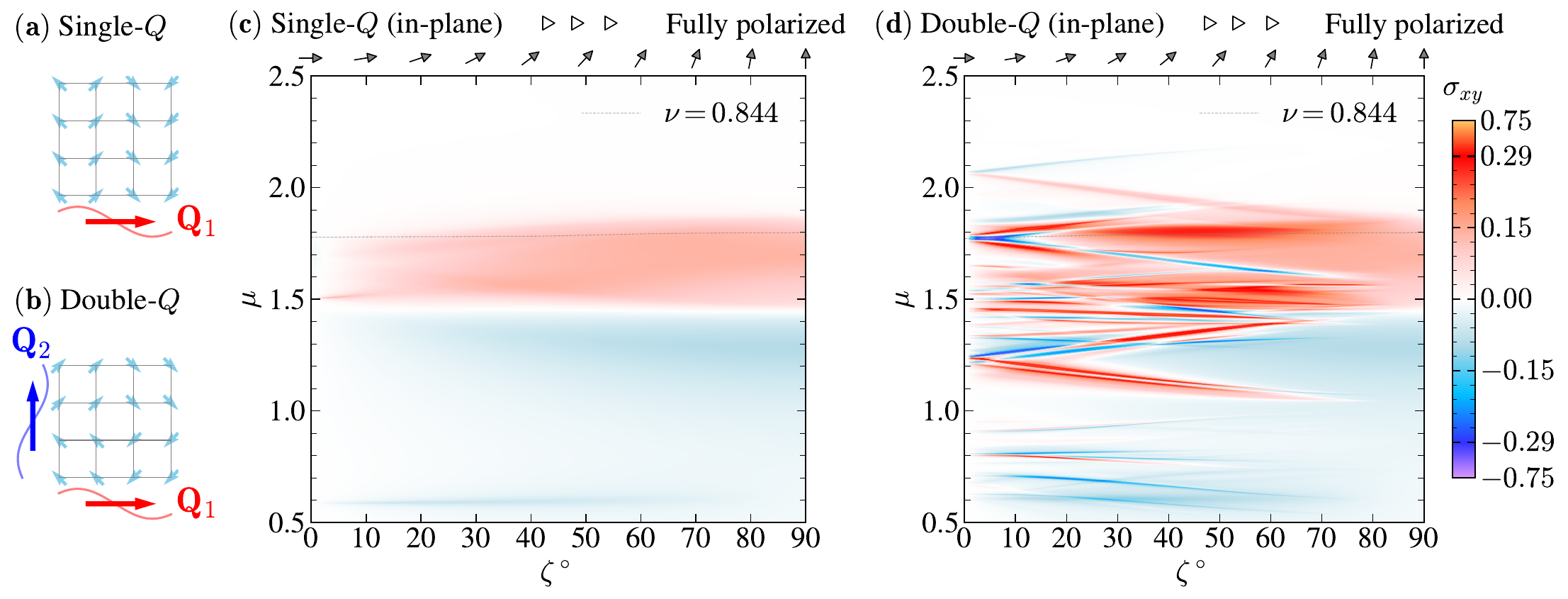}
    \caption{
    Panels (a) and (b) show the 1$Q$ and 2$Q$ spin configurations, respectively. 
    Panels (c) and (d) display the anomalous Hall conductivity $\sigma_{xy}$ for the 1$Q$ and 2$Q$ states as functions of the chemical potential $\mu$ and the canting angle $\zeta$ of the magnetic moments.
    Here, $\zeta = 0^{\circ}$ corresponds to a coplanar spin configuration, while $\zeta = 90^{\circ}$ represents a fully polarized state.
    The same color scale is applied to panels (c) and (d). 
    The dashed line marks the chemical potential corresponding to the filling ratio $\nu = 0.844$.
    }
    \label{fig:SIGMA_heatmap_upper}
\end{figure}

Figures~\ref{fig:SIGMA_heatmap_upper}(a) and \ref{fig:SIGMA_heatmap_upper}(b) show the spin textures of the 1$Q$ and 2$Q$ systems at $\zeta = 0^\circ$, respectively. 
Figures~\ref{fig:SIGMA_heatmap_upper}(c) and \ref{fig:SIGMA_heatmap_upper}(d) present the corresponding anomalous Hall conductivity for the 1$Q$ and 2$Q$ states. 
This figure corresponds to Fig.~1 of the main text, with the chemical-potential range shifted from $[-1.5, -0.5]$ to $[0.5, 2.5]$. 
The dashed lines in panels (c) and (d) indicate the chemical potential associated with the filling ratio $\nu = 0.844$. 
The values of $\sigma_{xy}$ along these dashed lines are used in Fig.~3 of the main text.

\subsection{Berry-curvature hot spots responsible for the enhancement of the anomalous Hall conductivity}
\begin{figure}
    \centering
    \includegraphics[width=\linewidth]{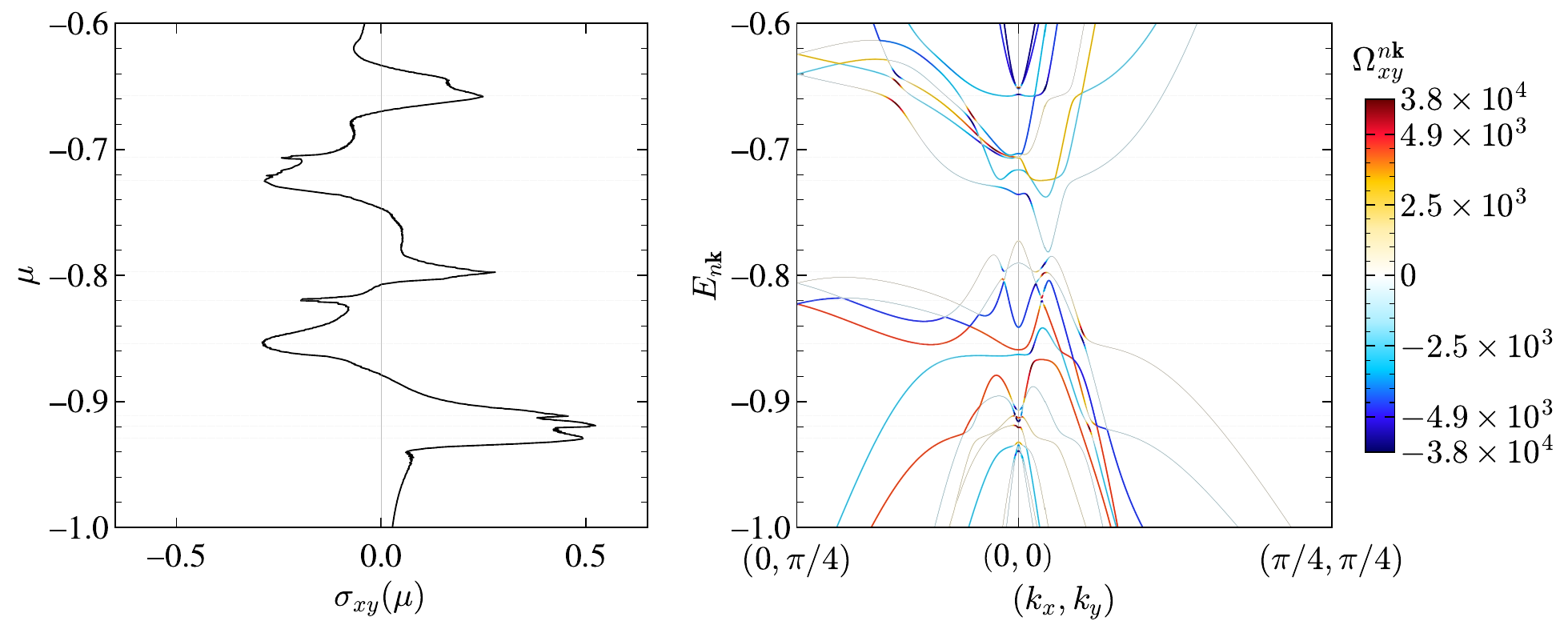}
    \caption{
    The left panel shows the anomalous Hall conductivity $\sigma_{xy}$ as a function of the chemical potential $\mu$. 
    The right panel displays the energy band structure along high-symmetry paths, colored by the Berry curvature $\Omega_{xy}$. 
    }\label{fig:SIGMA_vs_Omega}
\end{figure}

Figure~\ref{fig:SIGMA_vs_Omega} shows the anomalous Hall conductivity $\sigma_{xy}$ as a function of the chemical potential $\mu$ (left panel) and the corresponding Berry curvature $\Omega_{xy}$ mapped onto the energy bands along high-symmetry lines (right panel). 
One can confirm that the peaks in $\sigma_{xy}$ around $\mu = -0.8$ in the left panel originate from the large Berry-curvature hot spots near the $\Gamma$ point, as revealed in the right panel. 
Such a sharp enhancement of $\Omega_{xy}$ in this region emerges only when the double-$Q$ (2$Q$) spin configuration is considered.

\clearpage
\subsection{Generality of the enhancement of anomalous Hall conductivity in 2$Q$ state}

\begin{figure}
    \centering
    \includegraphics[width=1\linewidth]{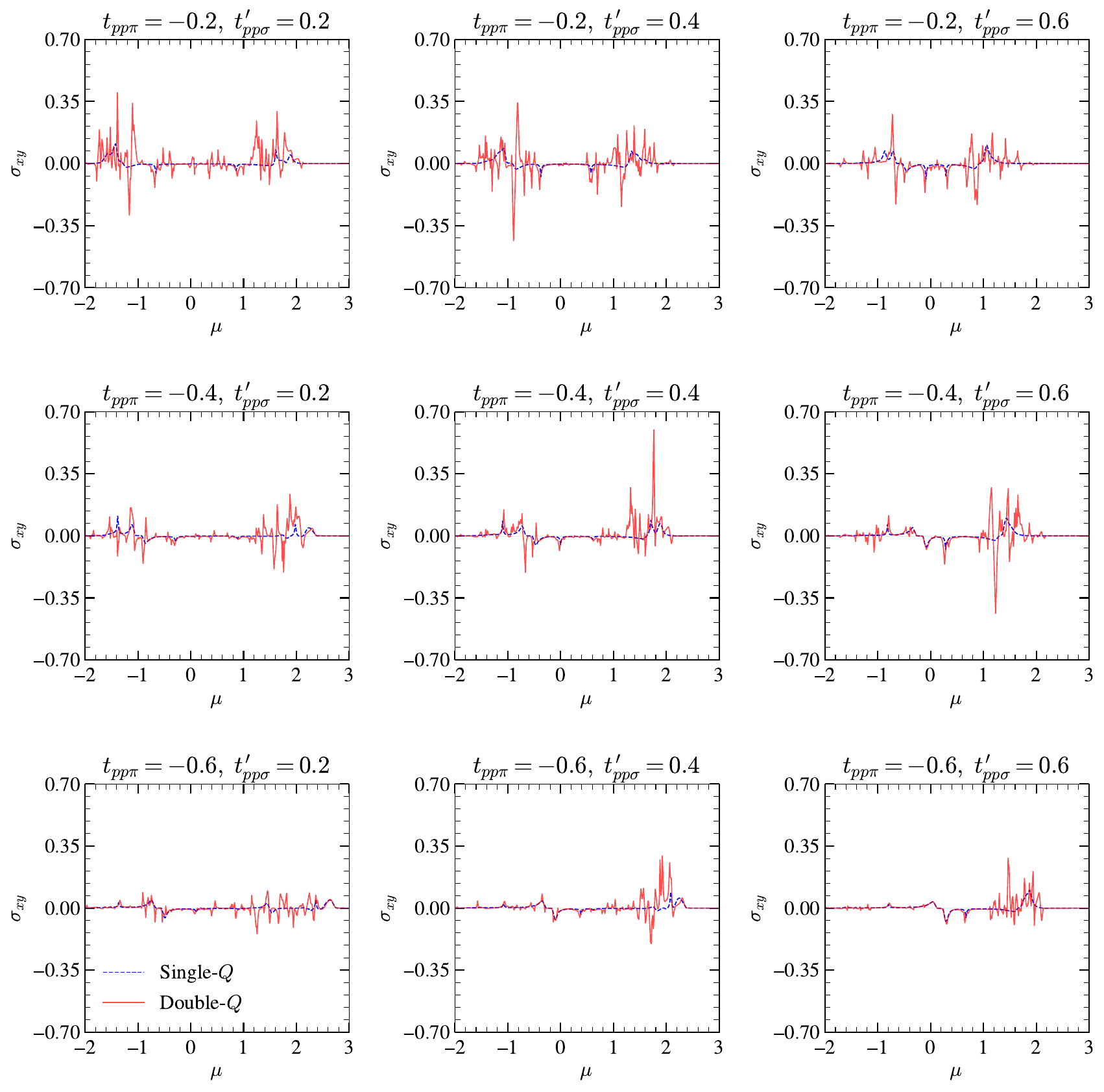}
    \caption{
    Anomalous Hall conductivity $\sigma_{xy}$ as a function of the chemical potential $\mu$ in the 1$Q$ (blue dashed curve) and 2$Q$ (red solid curve) states.
    The hopping parameters are varied as indicated in the panel titles: 
    specifically, $t_{pp\pi}$ is varied as $-0.2,\,-0.4,\,-0.6$ along the rows, while $t_{pp\sigma}'$ is varied as $0.2,\,0.4,\,0.6$ along the columns.
    The parameters $t_{pp\sigma} = 1$ and $t_{pp\pi}' = -0.1$ are fixed throughout.
    }\label{fig:various_hopping}
\end{figure}

Generality of the result is examined by varying the hopping parameters and the basis magnetic modulation vectors.  
First, we analyze the dependence on the hopping parameters.  
Figure~\ref{fig:various_hopping} shows the anomalous Hall conductivity $\sigma_{xy}$ as a function of the chemical potential $\mu$.  
The peak positions of $\sigma_{xy}$ shift as the hopping parameters are varied.  
However, the enhancement of $\sigma_{xy}$ in the 2$Q$ state appears robust across all parameter sets.  
The single-$Q$ (1$Q$) state consistently exhibits a smaller magnitude compared with the corresponding 2$Q$ state.

\begin{figure}
    \centering
    \includegraphics[width=0.5\linewidth]{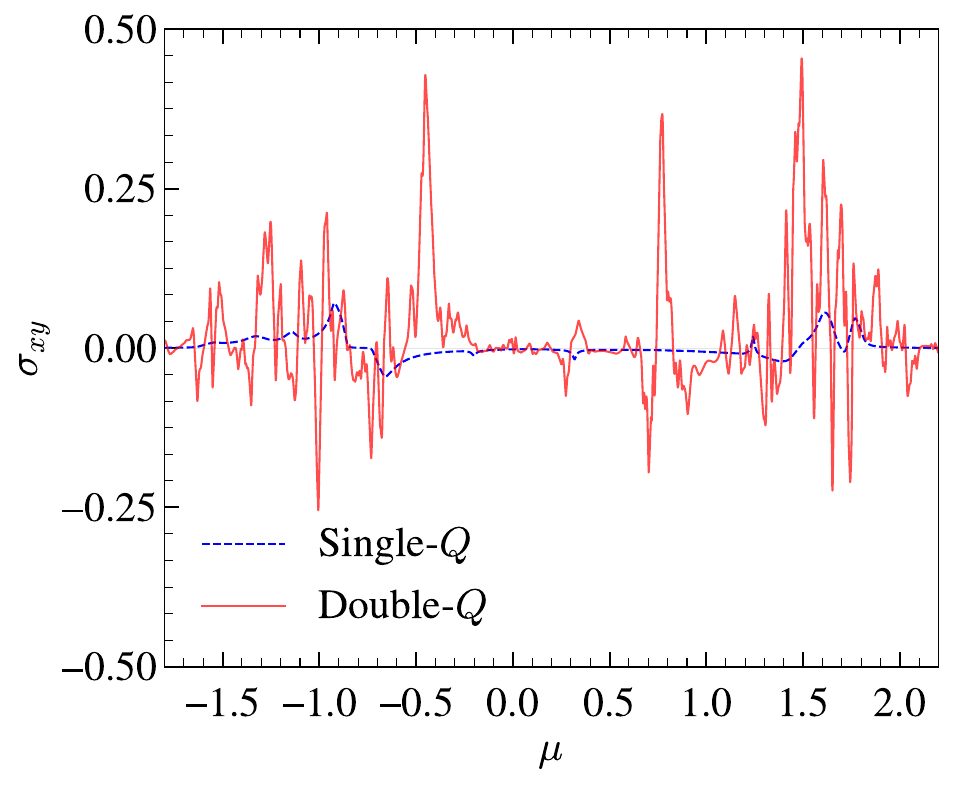}
    \caption{
    Anomalous Hall conductivity $\sigma_{xy}$ as a function of the chemical potential $\mu$ for the 2$Q$ state (red solid curves) and the 1$Q$ state (blue dashed curves).  
    The magnetic modulation vector for the 1$Q$ state is $\bm{Q}_{1} = (3\pi/4,\,0)$, while the 2$Q$ state consists of $\bm{Q}_{1} = (3\pi/4,\,0)$ and $\bm{Q}_{2} = (0,\,3\pi/4)$.
    }\label{fig:SIGMA_64site}
\end{figure}

Secondly, we change the magnetic modulation vectors from $\bm{Q}_1 = (\pi/2,0)$ and $\bm{Q}_2 = (0,\pi/2)$ to $(3\pi/4,0)$ and $(0,3\pi/4)$. 
The localized spin configurations retain the same functional form as in the main text, but are constructed using these new modulation vectors.
Figure~\ref{fig:SIGMA_64site} shows $\sigma_{xy}$ calculated under these modified spin textures.
The new 2$Q$ state still exhibits a consistently larger anomalous Hall response than the new 1$Q$ state.
This result demonstrates that the enhancement of $\sigma_{xy}$ in 2$Q$ spin textures persists even when the magnetic modulation period is changed, indicating that the effect is generic to a broad class of multi-$Q$ states.

\clearpage
\subsection{Parameter dependence of the Berry curvature in the 1$Q$ state}
\begin{figure}[t]
    \centering
    \includegraphics[width=0.6\linewidth]{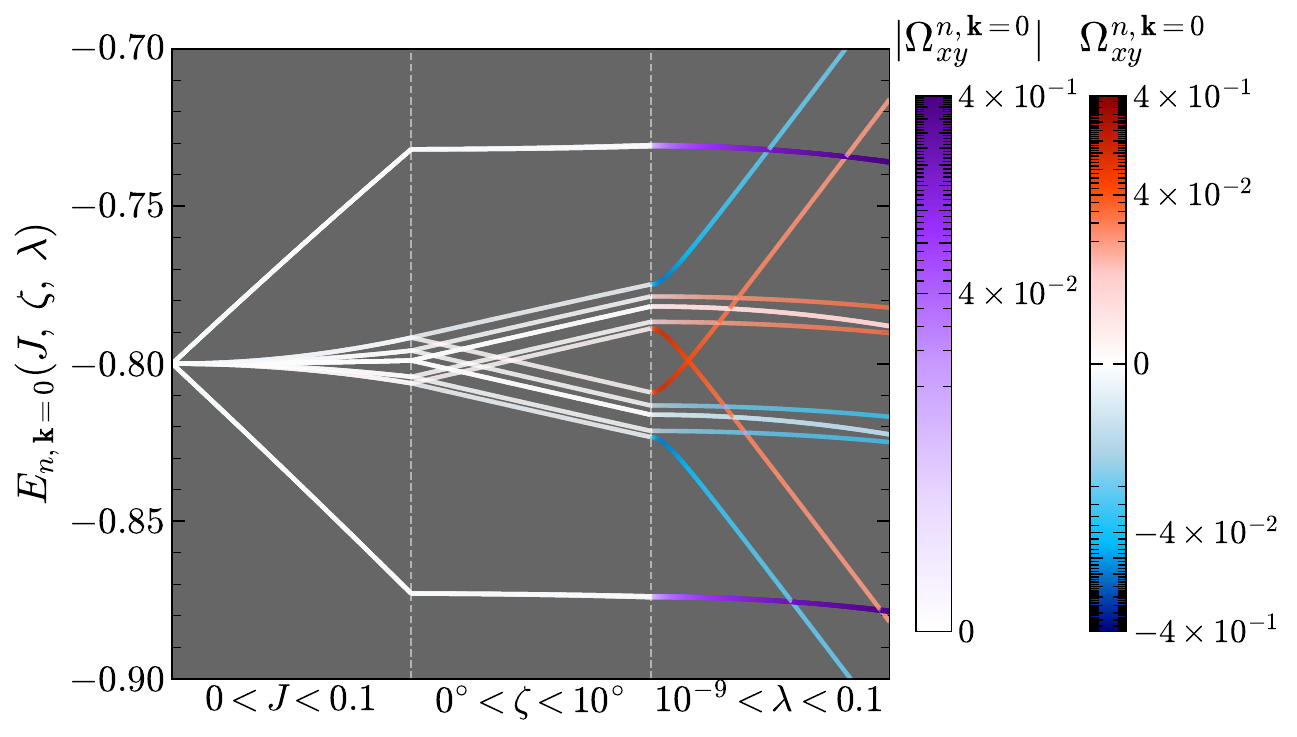}
    \caption{
    Berry curvature of the energy eigenstates at the $\Gamma$ point for the 1$Q$ magnetic configuration.
    The vertical axis denotes the eigenenergy, and the horizontal axis represents the trajectory in the model-parameter space $(J, \zeta, \lambda)$, where $J$ is the Kondo coupling constant, $\zeta$ is the out-of-plane magnetic angle, and $\lambda$ is the SOC constant.
    Along the parameter path, only one parameter is varied at a time while the others are held fixed:
    $(J,\zeta,\lambda): (0,0,10^{-9}) \rightarrow (0.1,0,10^{-9}) \rightarrow (0.1,10,10^{-9}) \rightarrow (0.1,10,0.1)$.
    Violet shading indicates the absolute value in degenerate regions where contributions cancel, while the blue-to-red colormap is used where degeneracy is lifted.
    }\label{fig:Omega_1Q}
\end{figure}

Figure~\ref{fig:Omega_1Q} shows the Berry curvature of the energy eigenstates at the $\Gamma$ point in the 1$Q$ state as a function of the model parameters.
In contrast to the large Berry curvature generated in the 2$Q$ state discussed in the main text, the 1$Q$ state does not exhibit any Berry curvature until the SOC is turned on.
Even when SOC is introduced, the resulting Berry curvature remains small, with typical magnitudes below $10^{-1}$, in sharp contrast to the values on the order of $10^{4}$ that arise in the 2$Q$ state.

\clearpage
\subsection{Microscopic origin of the Berry-curvature enhancement in the 2$Q$ state}

\begin{figure}
    \centering
    \includegraphics[width=\linewidth]{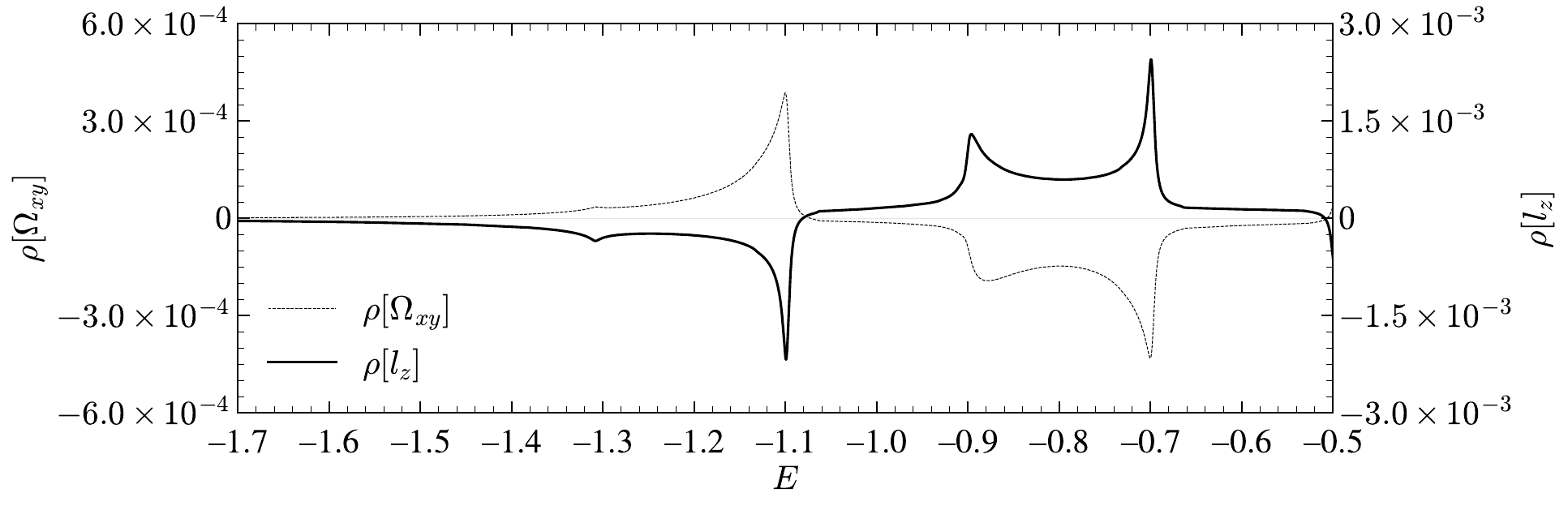}
    \caption{
    Density of Berry curvature $\rho[\Omega_{xy}]$ (dashed curve; left axis) and density of orbital angular momentum $\rho[l_{z}]$ (solid curve; right axis) as functions of the energy level $E$ in the fully polarized state.  
    }\label{fig:Omega_vs_lzQ_0Q}
\end{figure}

\begin{figure}
    \centering
    \includegraphics[width=\linewidth]{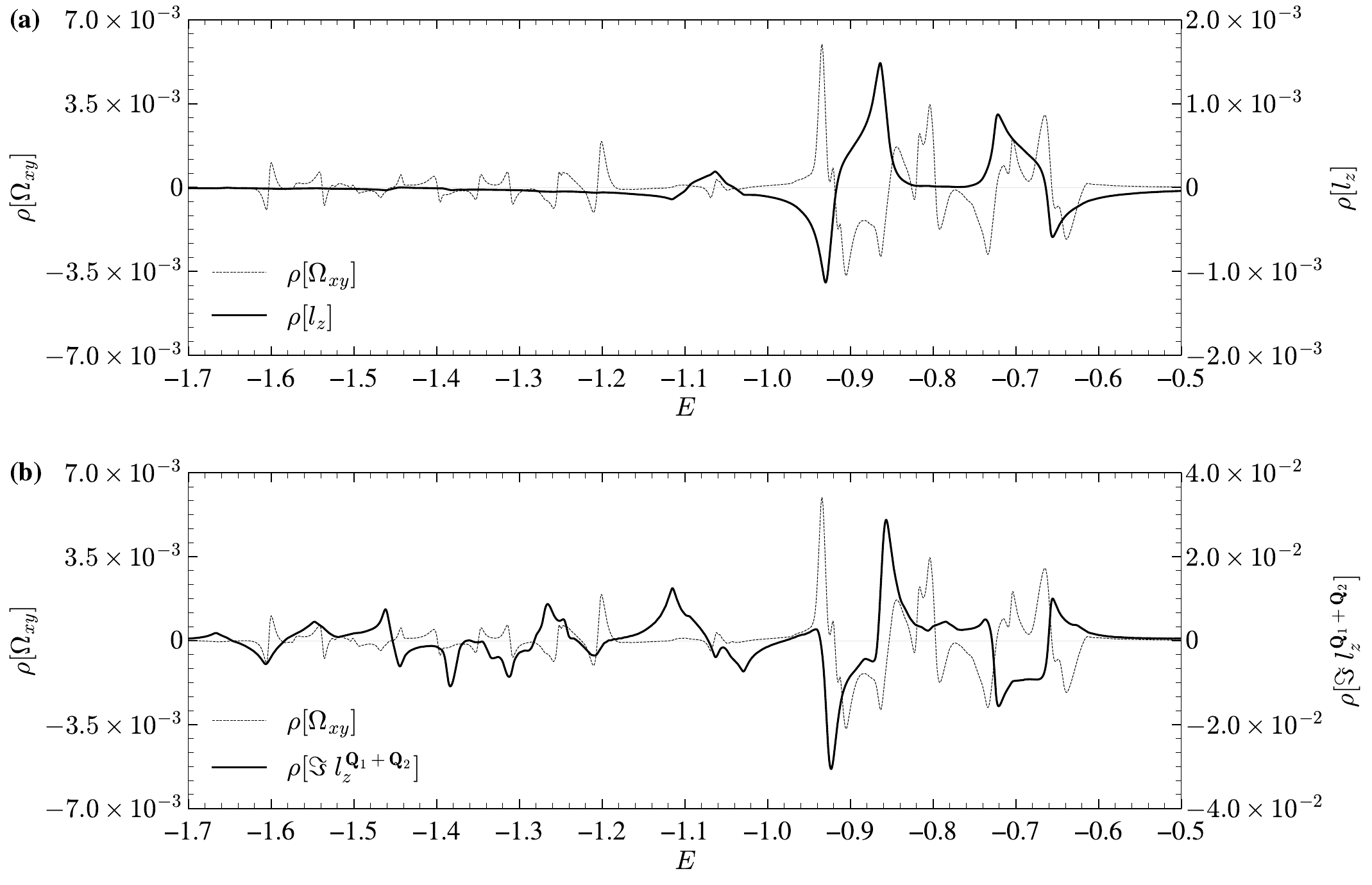}
    \caption{
    Panel (a) shows the density of Berry curvature $\rho[\Omega_{xy}]$ (dashed curve; left axis) and the density of orbital angular momentum $\rho[l_{z}]$ (solid curve; right axis) as functions of the energy level $E$ in the 2$Q$ state.  
    Panel (b) shows the density of Berry curvature $\rho[\Omega_{xy}]$ (dashed curve; left axis) together with the imaginary part of the $\bm{Q}_{1}+\bm{Q}_{2}$ component of the orbital-angular-momentum density $\rho[\Im\, l_{z}^{\bm{Q}_{1}+\bm{Q}_{2}}]$ (solid curve; right axis), plotted as functions of the energy level $E$.  
    }\label{fig:Omega_vs_lzQ_2Q}
\end{figure}

In the main text, we proposed that the enhancement of the Berry curvature in the 2$Q$ state arises from orbital hybridization generated by higher-order spin-scattering processes.  
Here we examine this mechanism in detail.  
A conventional viewpoint is that the anomalous Hall conductivity is closely tied to the orbital angular momentum~\cite{Xiao_2006_PhysRevLett.97.026603, Junren_2007_PhysRevLett.99.197202}.  
Figure~\ref{fig:Omega_vs_lzQ_0Q} compares the density of Berry curvature $\rho[\Omega_{xy}]$ with the density of orbital angular momentum $\rho[l_{z}]$ in the fully polarized state, where
\begin{equation}\label{eq:lz_density}
    \rho[X] = \frac{1}{N}\sum_{n\bm{k}} X_{n\bm{k}}\,\delta(E - E_{n\bm{k}}).
\end{equation}
In this case, the two quantities show a clear correlation.  
However, this correspondence breaks down in the presence of the 2$Q$ spin texture.  
As shown in Fig.~\ref{fig:Omega_vs_lzQ_2Q}(a), the peaks in $\rho[\Omega_{xy}]$ are not reproduced by $\rho[l_{z}]$; in particular, the prominent structures in the range $-1.7 < E < -1.1$ are entirely absent in the $\bm{Q}=\bm{0}$ contribution.
Thus, the uniform component of the hybridization alone cannot account for the large Berry curvature in the 2$Q$ state.

This leads to our interpretation that the enhanced Berry curvature in the 2$Q$ state originates from orbital hybridization at finite wave-vector modulation, induced by higher-order spin-scattering processes rather than by uniform orbital polarization.
To probe this mechanism, we evaluate the Fourier component of $l_{z}$ at $\bm{q}=\bm{Q}_{1}+\bm{Q}_{2}$, which represents the portion of the orbital hybridization generated by higher-order spin-scattering processes at this finite wave vector.
This specific choice of momentum follows directly from the momentum conservation encoded in the expression
\begin{align}
    \langle l_{z}^{\bm{q}} \rangle^{(2)}_{\text{con.}}
    = -\frac{J^2}{\beta N^2}
      \sum_{n,\bm{k},\bm{q}_1,\bm{q}_2}
      \operatorname{Tr}\!\big[
          l_{z}\,\cdot g_{\bm{k}}(i\omega_n)\, \cdot
          \sigma^{\mu}\, \cdot g_{\bm{k}+\bm{q}_1}(i\omega_n)\, \cdot
          \sigma^{\nu}\, \cdot g_{\bm{k}+\bm{q}_1+\bm{q}_2}(i\omega_n)
      \big]\,
      \delta_{\bm{q}_1+\bm{q}_2,\bm{q}} S_{\mu}^{\bm{q}_1} S_{\nu}^{\bm{q}_2},
\end{align}
where $g_{\bm{k}}(i\omega_{n}) = (i\omega_{n} - \mathcal{H}_{0}(\bm{k}))^{-1}$ is the noninteracting Green’s function.  
The trace is taken over both orbital and spin indices, and the Kronecker delta $\delta_{\bm{q}_1+\bm{q}_2,\bm{q}}$ enforces momentum conservation, ensuring that only connected processes contribute at second order. 
This expression gives the connected contribution to $l_{z}^{\bm{q}}$ induced by the second-order spin-scattering process. 
Because spin-scattering term carries finite Fourier components at $\bm{Q}_{1}$ and $\bm{Q}_{2}$ in the 2$Q$ magnetic state, this relation shows that $l_{z}$ acquires a finite component at $\bm{Q}_{1}+\bm{Q}_{2}$.

Figure~\ref{fig:Omega_vs_lzQ_2Q}(b) shows the density of Berry curvature $\rho[\Omega_{xy}]$ together with the density of the imaginary part of the orbital angular momentum $\rho[\Im\, l_{z}^{\bm{Q}_{1}+\bm{Q}_{2}}]$, plotted as functions of the energy level $E$.
Even though not every peak in $\rho[\Omega_{xy}]$ is captured by $\rho[l_{z}^{\bm{Q}_{1}+\bm{Q}_{2}}]$ alone, the structures in the range $-1.7 < E < -1.1$ are reproduced better than in Fig.~\ref{fig:Omega_vs_lzQ_2Q}(b) for the $\rho[l_{z}]$ case.
This supports our interpretation that the enhanced Berry curvature in the 2$Q$ state originates from orbital hybridization generated by higher-order spin-scattering processes involving the Kondo couplings $H_{K}^{\bm{Q}_{1}}$ and $H_{K}^{\bm{Q}_{2}}$.

\bibliographystyle{apsrev4-2}
\bibliography{ahe}